\documentclass{optica-article}

\journal{opticajournal} 

\articletype{Research Article}

\usepackage{lineno}

\usepackage{braket}
\usepackage{gensymb}

\usepackage{xcolor}

\usepackage{float}

\newcommand{\hide}[1]{}

\begin{document}

\title{Ultra-broadband, Low-loss Wavelength Combiners and Filters: Novel Designs and Experiments in Thin-film Lithium Niobate}

\author{
Robert Kwolek$^{1}$,
Parash Thapalia$^{2}$,
Aditya Tripathi$^{1}$,
Pooja Kulkarni$^{2}$, 
Jaber Balalhabashi$^{4}$,
Farzaneh Arab Juneghani$^{2}$,
Michael Bullock$^{1,5}$
Oanh Hoang Vo$^{1,6}$,
Sasan Fathpour$^{2,7}$,
and Rajveer Nehra$^{1,3,4,5^\ast}$
}

\address{
$^{1}$Department of Electrical and Computer Engineering, University of Massachusetts Amherst, Amherst, MA 01003, USA\\
$^{2}$CREOL, The College of Optics and Photonics, University of Central Florida, Orlando, FL 32816, USA\\
$^{3}$Department of Physics, University of Massachusetts Amherst, Amherst, MA 01003, USA\\
$^{4}$Materials Science and Engineering Graduate Program, University of Massachusetts Amherst, Amherst, MA 01003, USA\\
$^{5}$ Department of Computer Science, University of Massachusetts, Amherst, MA\\
$^{6}$ Department of Physics, Mount Holyoke College, South Hadley, MA 01075 USA\\
{$^{7}$ Department of Electrical and Computer Engineering, University of Central Florida, Orlando, FL 32816, USA}\\
$^*$Email: rajveernehra@umass.edu
}

\begin{abstract*} 
\\
Thin-film lithium niobate (TFLN) has emerged as a leading platform for large-scale programmable photonic circuits for quantum and classical applications. As circuits scale in complexity, low-loss routing of broadband pump and signal fields becomes essential. Here, we present closed-form analytical models and experimentally demonstrate compact, fast-quasi-adiabatic driving-optimized wavelength combiners and filters operating at the fundamental harmonic (FH, 1550 nm) and second-harmonic (SH, 775 nm) wavelengths. Our designs achieve ultra-low loss below $0.06~\mathrm{dB}$ across a $90~\mathrm{nm}$ bandwidth at FH, while maintaining extinction ratios exceeding $25~\mathrm{dB}$. At SH, the loss remains below $0.12~\mathrm{dB}$ over a $45~\mathrm{nm}$ bandwidth with extinction ratios greater than $19~\mathrm{dB}$. Devices fabricated on a $300$-nm TFLN platform exhibit added loss below $0.1~\mathrm{dB}$ across $1550$ -- $1600~\mathrm{nm}$, with minimum values of $0.04~\mathrm{dB}$ around $1580~\mathrm{nm}$ and $0.021~\mathrm{dB}$ at $775~\mathrm{nm}$. Combined with recent advances in on-chip quantum state generation,  low-loss interferometers, and detection, these results enable high-fidelity quantum photonic circuits on the TFLN platform.

\end{abstract*}

\section{Introduction}

Nanophotonics offers a practical route to build robust large-scale information processing and sensing systems by integrating light sources, programmable phase-stabilized interferometers, filters, and detectors on a compact chip~\cite{chrostowski2015silicon}. Silicon nitride and silica are mature nanophotonic platforms with ultra-low propagation loss and excellent fabrication processes, making them highly effective for many linear and nonlinear operations. However, their relatively weak third-order ($\chi^{(3)}$) nonlinearity typically relies on resonant enhancement, which introduces a trade-off between interaction strength and bandwidth, potentially limiting scalability in broadband frequency- and time-multiplexed photonic systems. Thin-film lithium niobate (TFLN) has recently emerged as a promising nanophotonic platform, combining strong second-order nonlinearity~\cite{Rao2019,Wang2018}, large electro-optic effect~\cite{Wang2018-EO}, quasi-phase-matching via periodic poling, and dispersion engineering to achieve single-pass traveling-wave nonlinear devices with orders of magnitude higher efficiencies and bandwidths compared to bulk LN devices~\cite{boes2023lithium, vazimali2022applications, Zhu2021,nehra2022few, ledezma2022intense_opa}.

As integrated photonic systems grow from function-specific devices to configurable large-scale circuits, efficient routing and filtering of optical fields across multiple wavelengths becomes increasingly important. In particular, nonlinear quantum photonic circuits often require simultaneous handling of pump and signal fields spanning octave-scale spectral separation, such as the telecom-band fundamental and its second harmonic. Achieving broadband wavelength multiplexing with low insertion loss and high extinction ratio is therefore critical for preserving quantum coherence and minimizing loss-induced degradation in nonlinear and quantum processes. However, conventional directional couplers and wavelength-selective elements often face trade-offs between footprint, bandwidth, and loss.

Conventional approaches typically employ adiabatic directional couplers due to their large fabrication tolerance and broadband operation~\cite{Ding:13, Xing:13, nehra2022few}. In these devices, the waveguide supermodes evolve adiabatically at the fundamental harmonic (FH) while remaining weakly coupled at the second harmonic (SH). Achieving this dual condition generally requires long interaction lengths, which increases propagation loss. To reduce the device footprint, fast quasi-adiabatic driving (FAQUAD) methods constrain the maximum adiabatic coefficient during the evolution of the waveguide parameters~\cite{Sun:09, FAQUAD}. Previous implementations have largely focused on optimizing the dominant constant-gap interaction region~\cite{Sun:09, Hung:19} (Region I, Fig.~\ref{fig:fig_1}a). However, the coupler transition regions can strongly influence overall device performance~\cite{Ng:19, Taras01012021}, and imperfections in these sections can degrade extinction ratios, increasing losses that are particularly detrimental for quantum photonic applications.

In this work, we extend conventional constant-gap FAQUAD approaches to the full device evolution and present closed-form analytical models for wavelength combiners and filters operating simultaneously at FH (1550~nm) and the SH (775~nm). We propose and experimentally demonstrate an ultra-broadband, low-loss quasi-adiabatic coupler with an arbitrary splitting ratio. The optimized designs exhibit total losses below 0.06 dB over 90 nm at the fundamental harmonic (FH) and 0.12 dB over 45 nm at the second harmonic (SH), representing an order-of-magnitude improvement in extinction-ratio-limited loss and bandwidth compared to conventional adiabatic designs. The devices are implemented on a 300-nm-thick thin-film lithium niobate platform with a waveguide width of $\sim 1.2~\text{\textmu m}$, a sidewall angle of $65^\circ$, and an etch depth of $\sim 100$ nm. Experimentally, we measure a total added loss $< 0.1$ dB across the 1550 -- 1600 nm wavelength range, with optimal performance of $0.04 \pm 0.02$ dB at the FH and $0.021 \pm 0.003$ dB at the SH.

\section{Proposed Design and Analysis Framework}
We follow the analysis described in Ref.~\cite{Ng:19}. Two parallel waveguides with weak interaction can be formalized with coupled mode equations, with interacting mode amplitudes $F_1$ and $F_2$\\
\begin{equation}
    \frac{d}{dz}
    \begin{bmatrix}
        F_1\\
        F_2
    \end{bmatrix}
    =
    -i
    \begin{bmatrix}
        -\beta_1&\kappa\\
        \kappa&-\beta_2
    \end{bmatrix}
    \begin{bmatrix}
        F_1\\
        F_2
    \end{bmatrix}, 
\end{equation}
where $\beta_1$ and $\beta_2$ are isolated propagation constants, and $\kappa$ is the coupling coefficient. This can be made relative to the reference frame of the supermodes through $G_{1,2}=F_{1,2}e^{-i\frac{(\beta_1+\beta_2)}{2}z}$, leading to\\
\begin{equation}
    \frac{d}{dz}
    \begin{bmatrix}
        G_1\\
        G_2
    \end{bmatrix}
    =
    -i
    \begin{bmatrix}
        -\Delta\beta&\kappa\\
        \kappa&\Delta\beta
    \end{bmatrix}
    \begin{bmatrix}
        G_1\\
        G_2
    \end{bmatrix},
\end{equation}
where $\Delta\beta=\frac{\beta_1-\beta_2}{2}$. Using polar coordinates, one can define $\Gamma$ and $\chi$ as the coupling magnitude and angle. As a result, we have

\begin{equation}
    \frac{d}{dz}
    \begin{bmatrix}
        G_1\\
        G_2
    \end{bmatrix}
   =-i\Gamma
    \begin{bmatrix}
        -\cos(\chi)&\sin(\chi)\\
        \sin(\chi)&\cos(\chi)
    \end{bmatrix}
    \begin{bmatrix}
        G_1\\
        G_2
    \end{bmatrix}=-i
    \hat{H}
    \ket{G}.
\end{equation}

The eigenvalues of $\hat{H}$ are $\pm\Gamma$, from which the eigenvectors corresponding to the instantaneous supermodes are
\begin{equation}
    \ket{+}=\begin{bmatrix}
        \sin\left(\frac{\chi}{2}\right)\\
        \cos\left(\frac{\chi}{2}\right)
    \end{bmatrix},
    \ket{-}=\begin{bmatrix}
        \cos\left(\frac{\chi}{2}\right)\\
        -\sin\left(\frac{\chi}{2}\right)
    \end{bmatrix}.
\end{equation}

Thus, adiabatic transfer of the optical field from the bottom to the top port requires that the coupling angle $\chi$ be varied continuously from $0$ to $\pi$ while suppressing excitation to the other supermode. Under this condition, optical power remains entirely in the $\ket{+}$ eigenstate throughout device evolution. 
Expressing the system in the instantaneous supermode basis via the transformation $\ket{G} = w_{+}\ket{+} + w_{-}\ket{-}$, the coupled-mode dynamics can be written as

\begin{align}
    \frac{dw_+}{dz}&=-i\Gamma{}w_++\frac{1}{2}\frac{d\chi}{dz}w_-\\
    \frac{dw_-}{dz}&=-\frac{1}{2}\frac{d\chi}{dz}w_++i\Gamma{}w_-,
\end{align}

from which it is clear that the interaction strength between modes is governed by the ratio
$\left|\tfrac{1}{2}\chi'(z)\right|/\Gamma$.
This motivates us to define the adiabaticity parameter~\cite{louisell1955analysis}
\begin{equation}
    \eta(z)=\frac{1}{2\Gamma}\frac{d\chi}{dz}.
    \label{eq:eta}
\end{equation}

The FAQUAD approach is to let $\eta(z)$ be a constant $\eta\ll1$~\cite{FAQUAD}. To establish an analytical solution, we use a model to relate the coupling dynamics to waveguide parameters, namely the difference in top width (TW) between waveguides $\Delta{}$TW and the top gap between waveguides $g(z)$. Since $\Delta{}\beta$ is solely a function of isolated modes, it does not contain a gap dependence and is solely a function of $\Delta{}$TW. For the geometries considered here, this function is linear for small values of $\Delta{}$TW, though this assumption is not essential and the design readily generalizes to other models.

For weakly coupled identical waveguides, $\kappa(z) = \kappa_0e^{-\frac{g(z)}{g_0}}$. We can approximate $\kappa$ to be constant with respect to $\Delta{}$TW for small variations in top width and extend the 1D expression. We characterize the accuracy of our reduced model in Section A of our Supplemental Document, which compares our modeled values to simulated values within a parameter sweep conducted in Tidy3D's mode solver. 
The simplicity of our model—where each parameter depends on a single independent variable—allows a univariate mapping between the physical waveguide geometry and the resulting coupling dynamics in the subsequent analysis.

The primary fabrication-imposed constraint of our coupler is the minimum achievable gap, denoted $g_m$, which is $0.8~$\textmu$\mathrm{m}$ for the devices considered here. As such, an optimal geometry for maximizing modal interaction while minimizing device length would consist of an extended section with a constant gap $g = g_m$, centered at $z = 0$ and of length $l_m$. However, in such a configuration, driving the mixing angle $\chi$ close to its asymptotic values $0$ and $\pi$ requires the parameter $\Delta$TW to reach substantially larger values than desired. Terminating the adiabatic evolution before this point leads to non-adiabatic coupling between the supermodes, resulting in oscillatory behavior in the device response. 

To overcome this limitation, we adopt an alternative strategy in which the FAQUAD evolution continues while the waveguides are gradually separated. This allows $\chi$ to evolve much closer to its extreme values, while ensuring that the eventual deviation from the FAQUAD condition occurs at a sufficiently large gap $g_c$, where residual modal interaction is negligible. An ideal separation curve would be Euler bends, commonly used for their adiabatic curvature evolution, which are described mathematically as:

\begin{equation}
    \frac{d\theta}{ds} = 2 a^2 s
    \quad \Rightarrow \quad
    \theta(s) = (a s)^2 + \theta_0 ,
\end{equation}
where $s$ denotes arc length, $\theta(s)$ is the local tangential angle, and $\theta_0$ is the initial angle. For small slopes, the curvature can be approximated paraxially as $\tfrac{1}{2}\, d^2 g / dz^2$, and the arc length coordinate as $s \simeq z - l_m/2$ (for our symmetric geometry, $y = g/2$). Under these assumptions, the curvature relation yields a cubic approximation for the gap profile
\begin{equation}
    \frac{1}{2}\frac{d^2 g}{dz^2}
    = 2 a^2 \left(z - \frac{l_m}{2}\right)
    \quad \Rightarrow \quad
    g_e(z)
    = \frac{2}{3} a^2 \left(z - \frac{l_m}{2}\right)^3 + g_m ,
\end{equation}
where the integration constants are chosen to ensure maximal continuity at $z = l_m/2$.

\begin{equation}
\kappa(z)=\kappa_0 \exp\!\left[-\frac{g(z)}{g_0}\right]
=
\begin{cases}
\kappa_m, 
& |z|\le \dfrac{l_m}{2},\\[6pt]
\kappa_m \exp\!\left[-\left(\dfrac{|z|-\dfrac{l_m}{2}}{z_0}\right)^{3}\right],
& |z|> \dfrac{l_m}{2},
\end{cases}
\label{eq:kappa_z}
\end{equation}
where $\kappa_m=\kappa_0e^{-\frac{g_m}{g_0}}$ and $z_0 = \left(\frac{3 g_0}{2 a^2}\right)^{1/3}$.

For an arbitrary coupling profile $\kappa(z)$, the mixing angle $\chi(z)$ corresponding to a constant adiabaticity parameter $\eta$ can be obtained using Eq.~\eqref{eq:eta} together with the relation $\kappa = \Gamma \sin\chi$. This yields
\begin{equation}
\cos\!\bigl(\chi(z)\bigr)
=
-2\eta \int \kappa(z)\,dz
\label{eq:chi_int}
=
\begin{cases}
-2\eta\,\kappa_m\, z,
& |z|\le \dfrac{l_m}{2}, \\[8pt]
-2\eta\,\kappa_m\,\mathrm{sgn}(z)
\left[
\displaystyle
\int_{0}^{|z|-\frac{l_m}{2}}
\exp\!\left(-\left(\dfrac{z'}{z_0}\right)^3\right)\,dz'
+
\dfrac{l_m}{2}
\right],
& |z|> \dfrac{l_m}{2}.
\end{cases}
\end{equation}

The integration constants are fixed by enforcing $\chi(0)=\pi/2$ and continuity at $z=\pm l_m/2$. Once $\chi(z)$ is known, the propagation-constant mismatch follows directly as $\Delta\beta(z)=\kappa(z)\cot\chi(z)$, enabling the extraction of the required taper profile $\Delta$TW(z) within any chosen parametric model. In addition, the adiabaticity parameter $\eta$ may be related to the desired asymptotic behavior of the device by imposing the condition $\chi(z\!\to\!\infty)=\pi$. Enforcing this constraint yields
\begin{equation}
    \eta
    =
    \frac{1}{
        \kappa_m
        \left(
            l_m
            +
            \frac{2}{3}\,
            {\bf\Gamma}\!\left(\tfrac{1}{3}\right)
            z_0
        \right)
    },
    \label{eq:e_t(l)}
\end{equation}
where ${\bf\Gamma}$ represents the Gamma function. The quantity
\(
l_m + \tfrac{2}{3}{\bf\Gamma}\!\left(\tfrac{1}{3}\right) z_0
\)
may be interpreted as an effective interaction length of the device. With this formulation, the complete coupler geometry is fully specified by the curve parameters $g_m$, $l_m$, and $a$, together with the chosen coupling model parameters. 
We emphasize that the FAQUAD protocols are typically implemented numerically by prescribing a constant adiabaticity parameter and solving for the corresponding control profiles. 

Here, we show that for the coupling envelopes of interest, the asymptotic boundary condition $\chi(z\!\to\!\infty)=\pi$ yields a closed-form expression for $\eta$ that directly links the FAQUAD parameter to device geometry and fabrication constraints. Notably, the asymptotic end behavior can be engineered to approach any coupling angle. This flexibility enables the realization of FAQUAD couplers with arbitrary power-splitting ratios (e.g., 3 dB), as detailed in Section B of our Supplemental Document. 

\begin{figure}
    \centering
    \includegraphics[width=1\linewidth]{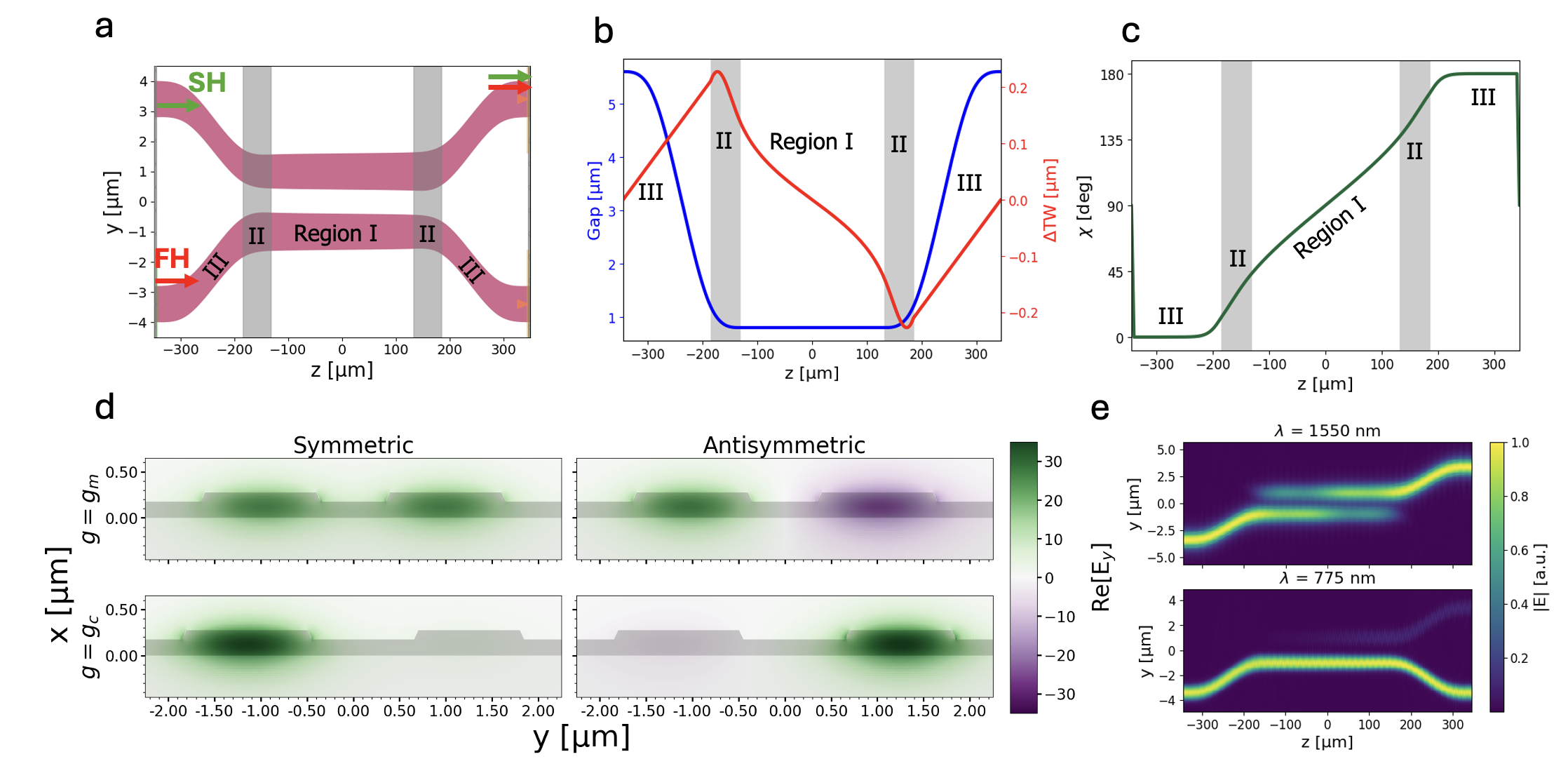}
    \caption{FAQUAD-optimized adiabatic directional coupler: design methodology and simulated performance. (a) Top-view geometry of the coupler showing the engineered longitudinal variation. Region I: Constant-gap section enabling controlled adiabatic supermode evolution. Region II: Cubic bend section designed to preserve adiabaticity during waveguide separation. Region III: Euler bends for low-loss routing and curvature-optimized transitions. (b) Spatial evolution of the waveguide top-width difference and inter-waveguide top gap. (c)  Prescribed FAQUAD coupling angle $\chi(z)$ ensuring controlled adiabatic evolution between supermodes. (d) Mode profiles of the symmetric and antisymmetric supermodes at the device's midpoint ($g_m = 800$~nm) and at the end of FAQUAD evolution ($g_c = 1200$~nm) where coupling is negligible. (f) Simulated nominal performance at FH and SH mode propagation, showing the high coupling for FH and high extinction for SH, as desired.}
    \label{fig:fig_1}
\end{figure}

We now turn our attention to simulations performed using Tidy3D - a commercial service with Finite Difference Time Domain
(FDTD) and mode solvers~\cite{tidy3d_flexcompute}. 
Figure~\ref{fig:fig_1}a depicts the coupler layout, comprising three distinct regions (I–III) that govern the longitudinal evolution of the supermodes. Region I corresponds to a constant-gap straight section in which the top and bottom waveguides are adiabatically tapered to initiate controlled mode evolution. Region II introduces the proposed cubic bends during the separation stage, carefully engineered to preserve adiabaticity and suppress unwanted mode excitation as the coupling strength decreases. Region III employs Euler bends to minimize propagation loss; this section begins once the waveguides are sufficiently separated such that residual coupling beyond the end of Region II is negligible. When pumped from the left with SH (green arrow) in the top port and FH (red arrow) in the bottom port, the device combines the two wavelengths at the top port, acting as a wavelength combiner. Conversely, if SH and FH are launched into the same input port, the coupler separates them into different waveguides, enabling their use as filters for quantum photonic experiments.

Figure~\ref{fig:fig_1}b shows the longitudinal evolution of the waveguide gap, $g(z)$ (blue, left axis), and the top-width difference, $\Delta\mathrm{TW}(z)$ (red, right axis). The central interaction region maintains a constant, fabrication-limited minimum gap $g_m$ to maximize the modal coupling. On either side, smoothly engineered S-bend separations adiabatically reduce the coupling strength. The shaded regions denote the cubic transitions that connect the constant-gap interaction section to the Euler-bend separation regions, terminating at a gap $g_c$. In Region III, $\Delta\mathrm{TW}$ is linearly tapered to zero, enhancing fabrication tolerance and bandwidth by reducing the local adiabaticity parameter $\eta$. This comes at the cost of a small additional insertion loss as a result of the slight modal discontinuity.

Figure~\ref{fig:fig_1}c shows the corresponding evolution of the coupling (mixing) angle $\chi(z)$, which increases monotonically from $0$ to $\pi$ under the constant-adiabaticity (FAQUAD) condition, ensuring complete and non-oscillatory mode transfer. Near the end of the device, though $\chi$ rapidly approaches $\pi/2$ as both waveguide widths become identical, the gap is sufficient to dissuade directional coupling.

Figure~\ref{fig:fig_1}d shows the simulated symmetric and antisymmetric supermodes at the FH wavelength. The top panels correspond to the midpoint of the device, where coupling between the waveguides is strongest due to the symmetric structure and minimum waveguide gap. The bottom panels show the mode profiles at the end of FAQUAD evolution, where the sufficiently large gap and width difference make coupling negligible, evidenced by both supermodes predominantly residing in opposing waveguides.

Finally, Figure~\ref{fig:fig_1}e presents full-device propagation simulations at the FH (top) and SH (bottom) wavelengths. The fields evolve smoothly and adiabatically, without observable beating or radiation loss. These results confirm near-unity coupling efficiency at FH and high extinction at SH, as analyzed in detail in the following sections. \\

To validate the theoretical design, we perform full-wave FDTD simulations using Tidy3D~\cite{tidy3d_flexcompute}. The device geometry is defined through parameterized GDSFactory functions, directly importing our exact layout parameters into the simulation environment and ensuring one-to-one consistency between design and numerical modeling. Low-resolution simulations were first employed to efficiently explore the multidimensional parameter space. Subsequently, high-resolution FDTD simulations were performed for fine optimization and fabrication-tolerance sweeps, using the waveguide parameters defined above. The final design parameters were chosen as $l_m = 264~\mu\mathrm{m}$, $a = 2~\mathrm{mm}^{-1}$, and $g_c = 1.2~\mu\mathrm{m}$, corresponding to an adiabaticity parameter of $\eta \approx 0.189$.

\begin{figure}
    \centering
    \includegraphics[width=0.95\linewidth]{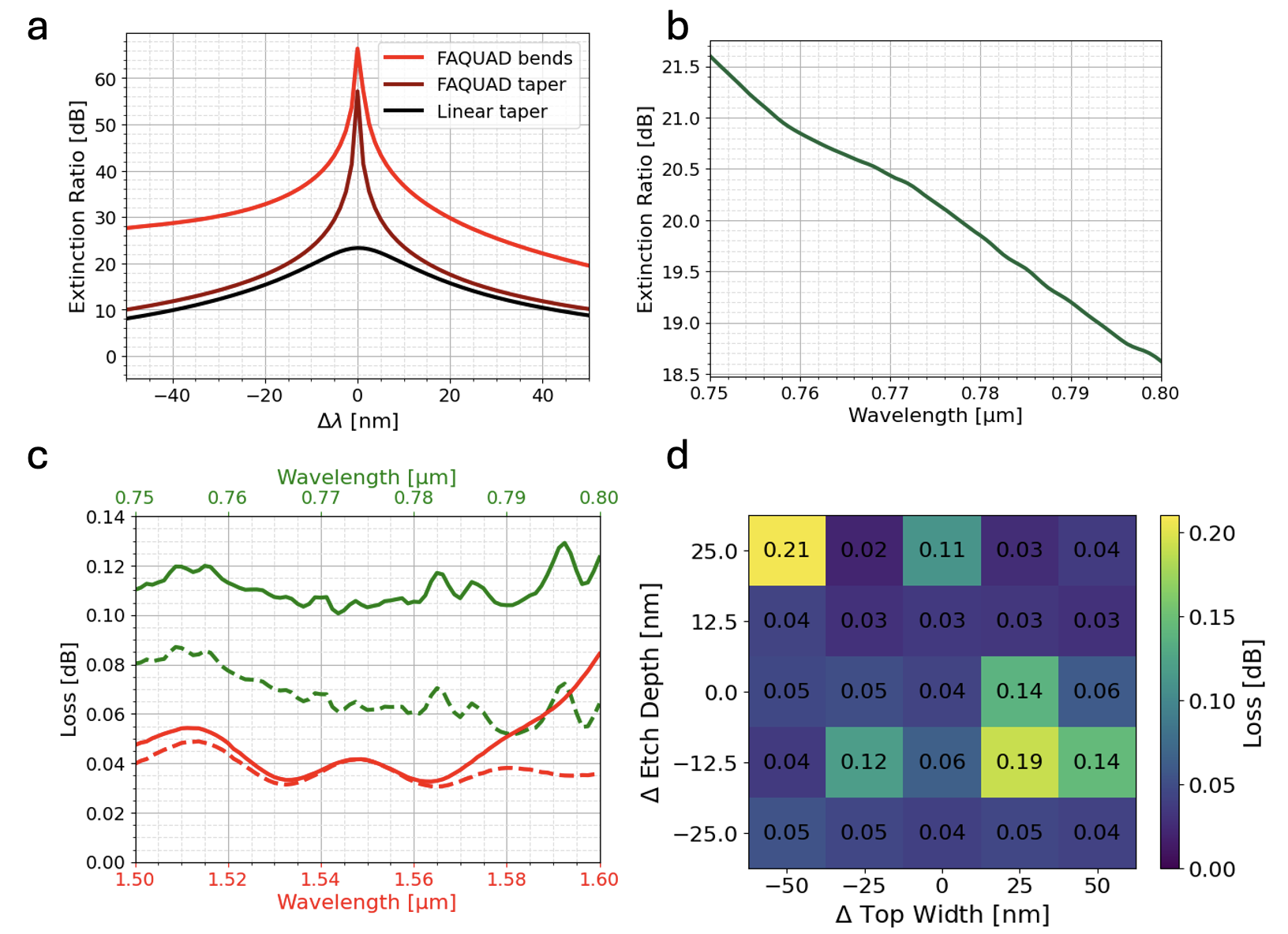}
    \caption{Simulated performance and fabrication-error tolerance. (a) Extinction Ratio at FH for our design [top] and others of the same length. (b) Extinction Ratio at SH. (c) Solid: Total loss at FH and SH. Dashed: Power lost from both outputs. (d) Total loss at FH over fabrication parameter sweep.}
    \label{fig:Tolerance}
\end{figure}

Figure~\ref{fig:Tolerance}a quantifies the FH extinction ratio, revealing a broadband response enabled by the global FAQUAD-engineered evolution (red curve). In contrast to designs that enforce constant adiabaticity only within the straight interaction region, the proposed design suppresses nonadiabatic mode coupling. As a result, the extinction exceeds $25~\mathrm{dB}$ over a range of $ 90~\mathrm{nm}$, indicating ultra-broadband performance and making it suitable for traveling-wave quantum circuits. At SH, shown in Figure~\ref{fig:Tolerance}b, the device operates in the weak-coupling regime, yielding an extinction ratio of more than $\sim 20~\mathrm{dB}$ at nominal wavelength and > 19 dB over 45 nm bandwidth. This weak interaction suppresses unintended power transfer while preserving spectral flatness. We emphasize that the present design is primarily optimized for the FH band, where the quantum signal typically resides and where minimal losses are most critical. The spectral response can be rebalanced to further enhance extinction at the SH wavelength; however, this would come at the expense of a slight performance degradation at FH, reflecting the inherent trade-off imposed by the coupled-mode dynamics and wavelength-dependent coupling strength. 

Figure~\ref{fig:Tolerance}c presents both propagation limited and total losses. The sold lines show total power lost from the desired port, while the dashed lines represent power lost from both outputs from scattering and radiation, which remain minimal due to adiabatic modal evolution and curvature-optimized Euler transitions. The difference between the two demonstrates extinction ratio limited losses, which is minimal across most of the FH bandwidth, and relatively constant across the SH bandwidth. The simulated total loss is below $0.06~\mathrm{dB}$ at FH over 90 nm, and below $0.12\mathrm{dB}$ at SH over 45 nm bandwidth. 

Finally, Figure~\ref{fig:Tolerance}d evaluates fabrication tolerance at FH by sweeping critical geometrical parameters. The total loss remains below $0.21~\mathrm{dB}$ across the explored deviations of 50 nm in the etching depth and 100 nm in the waveguide width, with most parameter variations producing losses $\le 0.05~\mathrm{dB}$. This robustness arises from the reduced sensitivity of the adiabatic transfer to small offsets in $\kappa$ and $\Delta\beta$, confirming the stability of the FAQUAD-based design against realistic fabrication errors.


\section{Experimental Results}

We fabricate our devices on a 300-nm x-cut TFLN wafer obtained from NanoLN. Prior to fabrication, ellipsometry measurements are performed across the wafer to identify regions with minimal variation in the thin-film thickness (\textasciitilde{}1.5nm RMSE along the propagation length). A chip is then diced from one of these uniform regions and used for device fabrication. The chip is first annealed at $520\,^{\circ}\mathrm{C}$ for $\sim$2~hours under an inert atmosphere \cite{shams2022reduced}. Waveguides are then patterned using hydrogen silsesquioxane (HSQ) resist with 100~kV electron-beam lithography. The patterned HSQ serves as an etch mask to transfer the structures into the TFLN layer via Ar ion etching in an ICP-RIE system, with the etch conditions optimized to achieve a rate of approximately 22~nm/min and an etch depth of $\sim$100~nm. After etching, the waveguides are cleaned in an RCA-1 solution to remove redeposited residues. Electrodes are subsequently patterned in ZEP520A resist and deposited using an electron-beam evaporator, consisting of a 20~nm chromium adhesion layer followed by 100~nm of gold. Finally, lift-off in N-methyl-2-pyrrolidone (NMP) is performed to define the electrode structures.

\begin{figure}
    \centering
    \includegraphics[width=1\linewidth]{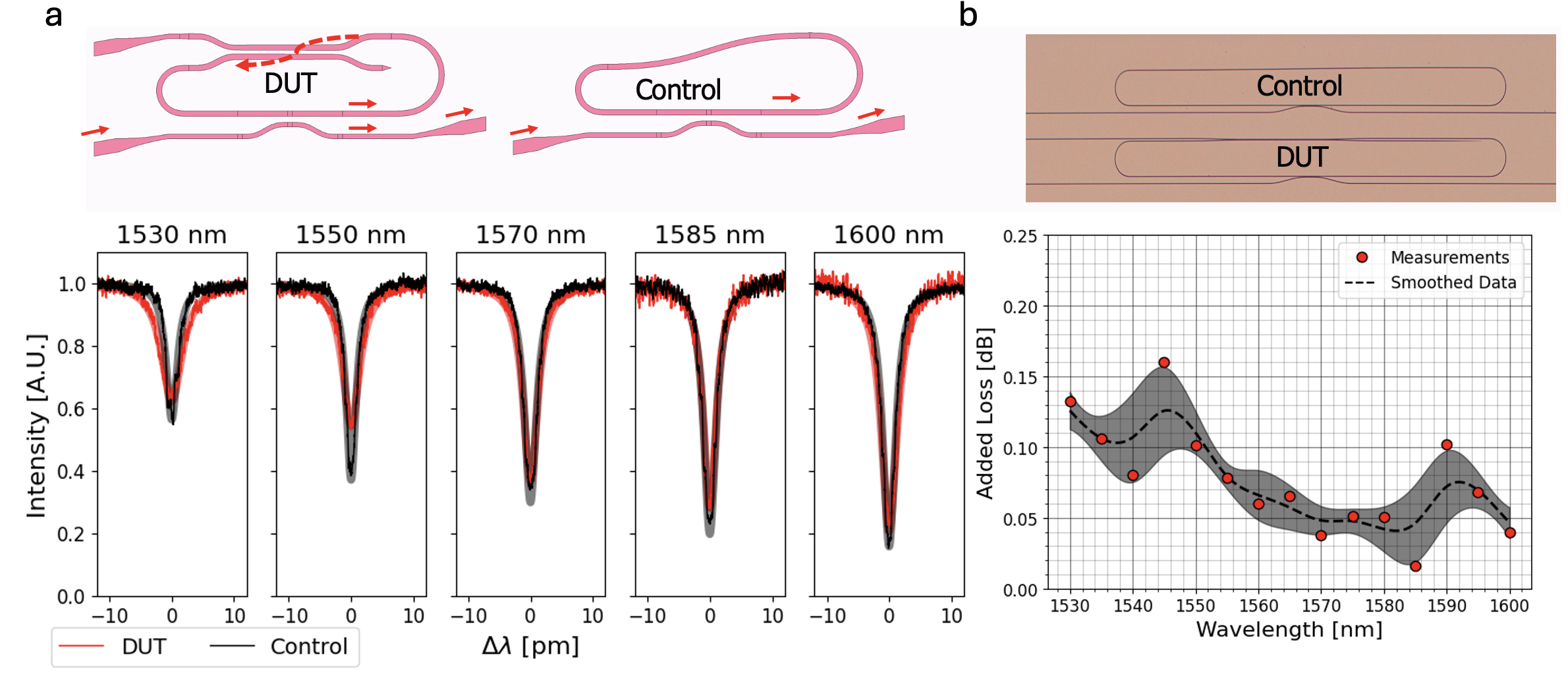}
    \caption{Experimental characterization. (a) Top: Measurement configuration used for FH characterization. Bottom: Representative resonance dips measured around the FH wavelength (1530–1600~nm) for the control resonator (black) and the resonator containing the DUT (red). The reduced resonance depth and increased linewidth in the DUT resonator indicate additional loss. (b) Top: Optical microscope image of the fabricated devices. Bottom: Extracted excess loss relative to the control resonator as a function of wavelength. Red points correspond to measured values, while the dashed curve shows the smoothed fit with the shaded region indicating the 95\% confidence interval.
}
\label{fig:Measurements}
\end{figure}

We evaluate the performance of our devices by integrating them into a resonator and comparing their transmission spectra with those of a control resonator without the component (Fig.~\ref{fig:Measurements}a, Top). A well-designed and fabricated filter will minimally degrade the resonator line width, indicating a small change in loaded $Q$ and thus low excess loss. Mathematically, we fit the power transfer function

\begin{equation}
    |H(\theta)|^2=
    \frac{L_p^2+L_c^2-2L_pL_c\cos(\theta)}
    {1+L_p^2L_c^2-2L_pL_c\cos(\theta)},
\end{equation}
where $L_p$ is the round-trip field transmission due to propagation loss, $L_c$ is the field self-coupling coefficient of the coupler, and
\begin{equation}
    \theta=\beta_{\mathrm{eff}}l+\theta_0,
\end{equation}
with $\beta_{\mathrm{eff}}$ as the effective propagation constant, $l$ being the round-trip cavity length, and $\theta_0$ as a constant phase offset. Because $|H(\theta)|^2$ is symmetric under interchange of $L_p$ and $L_c$, the two coefficients cannot be uniquely extracted from the transmission spectrum alone without additional information~\cite{yariv2000universal,bogaerts2012silicon}. We resolved this ambiguity by comparing a control resonator with a resonator that contained the device under test (DUT). Since both resonators employ identical couplers, the coupling coefficient $L_c$ is expected to be the same for both devices. We therefore identify the coefficient that remains constant between the two fits as $L_c$, while any variation is attributed to changes in the round-trip propagation coefficient $L_p$. The corresponding power transmission factors are $T_{p,c}=L_{p,c}^2$, and the associated losses in decibels are

\begin{equation}
    \mathcal{L}_{p,c}
    =-20\log_{10}(L_{p,c}).
\end{equation}

To isolate the excess loss introduced by the device under test (DUT), we subtract the round-trip propagation losses extracted from resonator measurements,
$\Delta \mathcal{L}_{\mathrm{DUT}} = \mathcal{L}^{\mathrm{DUT}}_{p} - \mathcal{L}^{\mathrm{control}}_{p}$. To account for the propagation loss associated with the DUT, we estimate the missing propagation loss contribution as
\begin{equation}
\mathcal{L}_{\mathrm{propagation}}^{\mathrm{DUT}} = \frac{l_{\mathrm{DUT}}}{l}\,\mathcal{L}^{\mathrm{control}}_{p},
\end{equation}
where $l_{\mathrm{DUT}}$ is the physical length of the DUT section and $l$ is the total round-trip cavity length of the control resonator. 

Figure~\ref{fig:Measurements} shows the measurements at the FH, where the DUT is incorporated into the resonator (Fig.~\ref{fig:Measurements}a, top). In this configuration, the coupling efficiency of the DUT modifies the round-trip loss and consequently affects the measured quality factor relative to the control resonator. Figure~\ref{fig:Measurements}a (bottom) shows representative broadband resonance dips measured from 1530 to 1600 nm. The red and black traces correspond to the resonator containing the DUT and the control resonator, respectively. As expected, the inclusion of the DUT reduces the resonance depth and broadens the linewidth, consistent with increased under-coupling relative to the control device. At longer wavelengths, the control and DUT resonators have nearly identical responses, indicating low added loss.

Figure~\ref{fig:Measurements}b summarizes the extracted excess loss as a function of wavelength, with 95\% confidence intervals shown for the smoothed fit. Remarkably, the added loss remains below 0.1~dB over a 50~nm bandwidth and reaches a minimum of $0.04 \pm 0.02$~dB near 1580~nm. These results show good agreement with the simulated predictions, aside from a small shift in the nominal wavelength. Accounting for the intrinsic propagation loss of the DUT section would contribute an additional $\sim$0.03~dB. The measured added loss (solid red points) was smoothed (dashed line) over wavelength with a simple low-pass binomial kernel, giving estimates and confidence intervals as shown. All resonators were characterized from 1530 to 1600 nm using a continuously tunable laser source Toptica CTL~1550 with absolute wavelength accuracy $<150$~pm, linewidth 0.3~kHz.

\begin{figure}[!h]
    \centering
    \includegraphics[width=0.55\linewidth]{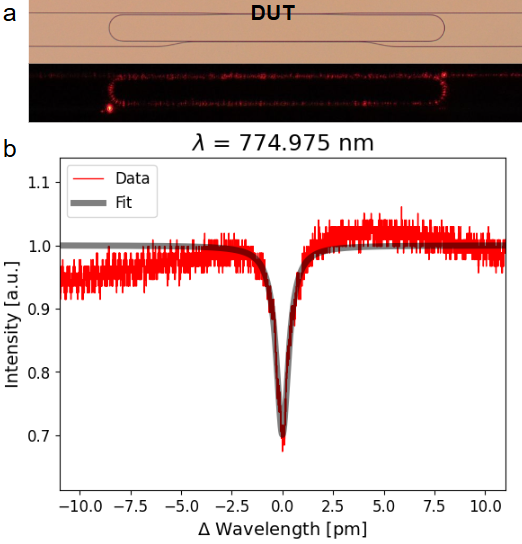}
    \caption{Characterization at the SH wavelength. (a) Top: Optical micrograph of the DUT where the FAQUAD-optimized coupler is used. Bottom: Micrograph image of the circuit in resonance when coupled from the DUT port. (b) Representative transmission resonance measured near $\lambda = 775$ nm. The red trace shows the measured transmission spectrum, while the gray curve corresponds to the fit obtained using the resonator transmission model.}
    \label{fig:fig775}
\end{figure}

For the SH measurements, the DUT is used directly as the coupler to a resonator (Fig.~\ref{fig:fig775}). In this configuration, the coupling loss is equivalent to the DUT excess loss, under the assumption that any additional insertion loss is comparable to that of straight propagation. This approach enables characterization without the need for a separate control resonator, with the primary ambiguity arising from the difficulty of distinguishing between round-trip propagation loss and coupling loss. In our case, this ambiguity can be resolved using prior knowledge of the propagation loss at the SH wavelength. Independent measurements indicate propagation losses of $\sim$1~dB/cm at 775~nm, whereas the alternative interpretation of the fit would imply much lower loss of $\sim$0.08~dB/cm. We therefore attribute the extracted loss predominantly to the coupling process. Although the resonator was originally designed to include an additional directional coupler to allow independent extraction of $L_p$, this coupler was nonfunctional due to fabrication error; however, the remaining ports were sufficient for the present characterization.

Figure~\ref{fig:fig775} shows a representative resonance near $\lambda_{\text{SH}} = 775$ nm.  Multiple resonances were measured in the vicinity of the SH band to statistically extract the coupling-induced excess loss. From these measurements, we determine an average coupling-induced loss of $0.021 \pm 0.003$ dB, confirming minimal parasitic coupling and high-quality mode isolation at the SH wavelength. The SH resonators were characterized near 775~nm using a narrowband SHG source. The asymmetry in the measured transmission originates from the wavelength-dependent output of the SHG source, which is optimized near 775~nm.

\section{Conclusion}

In summary, we have designed, fabricated, and experimentally characterized a compact wavelength combiner/filter based on fast quasi-adiabatic driving in thin-film lithium niobate. The devices enable efficient routing of FH and SH with ultra-low insertion loss and broadband operation. Resonator-based measurements show that the added loss remains below 0.1~dB over a 50~nm bandwidth with a minimum of $0.04 \pm 0.02$~dB near 1580~nm, in good agreement with numerical simulations. At the SH, coupling measurements indicate similarly low excess losses of $0.021 \pm 0.003$~dB, confirming the broadband and low-loss nature of the design. These results demonstrate that FAQUAD-engineered couplers with cubic bends provide an effective approach for compact and fabrication-tolerant wavelength routing in thin-film lithium niobate.  When combined with recently demonstrated traveling-wave quantum light sources, low-loss interferometers, and high bandwidth detectors, these devices form a key building block for scaling large-scale quantum photonic circuits.

\section{Funding}
This work was supported in part by the DARPA INSPIRED Program (D24AC00154-00). The views, opinions, and/or findings expressed are those of the author(s) and should not be interpreted as representing the official views or policies of the U.S. Government or any agency thereof. RN acknowledges the support of the College of Engineering, University of Massachusetts Amherst. The UCF contribution work is partially supported by NSF Industry University Cooperative Research Center (IUCRC) EPICA program.


\bibliography{sample}

\newpage
\appendix
\section*{Supplemental Document}

\section{Coupling Model Verification}

\begin{figure}[H]
    \centering
    \includegraphics[width=0.8\linewidth]{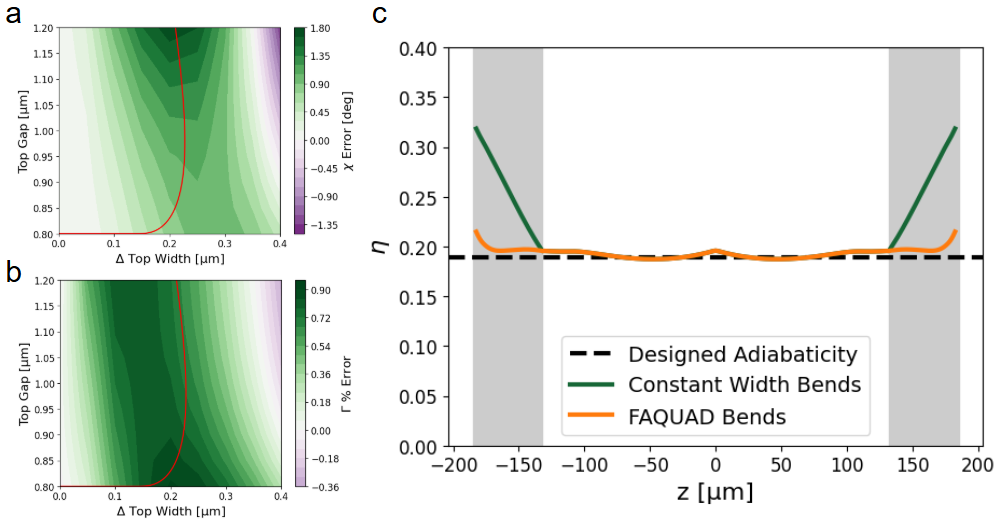}
    \caption{(a) Deviation of the implemented coupling angle $\chi$ from the target model (in degrees).
(b) Relative error of the coupling magnitude, $\Gamma$.
(c) Simulated adiabaticity relative to the design value in the FAQUAD sections of the coupler.}
    \label{fig:fig_1SI}
\end{figure}

Figures~\ref{fig:fig_1SI}a-b validate the precision of our reduced models, showing deviations below $2^\circ$ for the mixing angle $\chi$ and within $1\%$ for the coupling magnitude $\Gamma$, underscoring the robustness of the proposed design framework within the relevant parameter space. The red line indicates the curve traced by the proposed design, which is shown to perform well despite traversing some higher-error regions, suggesting that subsequent numerical optimization would further improve performance. Figure \ref{fig:fig_1}c uses simulated values of $\chi$ and $\Gamma$ to compare the true adiabaticity over length with the constant design value. Strong agreement in Sections I and II indicates high model fidelity, although more robust parameterizations would further reduce error. As expected, FAQUAD-evolved bends maintain a relatively constant adiabaticity, whereas constant-width bends exhibit severe deviations.

\section{Design Framework for Arbitrary Splitting Ratios through End Behavior}

One way to achieve an arbitrary splitting ratio is to let one side of the coupler asymptotically tend toward full coupling while the other end enforces the desired ratio. In contrast, letting both sides end on partial coupling would create oscillatory behavior along length due to the presence of both supermodes. As such, some of the previous convenient symmetry must be broken.

The splitting ratio being an arbitrary value motivates us to define $\chi_f=\lim_{z\rightarrow\infty}{\chi(z)}$, with $\sin^2(\chi_f/2)$ being the final power in the coupled port. Given that $g(z)$ will be an equivalent structure, we can let $\kappa{}(z)$ be identical to Eq. (10). This allows us to formulate $\cos(\chi(z))$ equivalently to Eq. (11), with the inclusion of an integration constant $C$ as $\chi(0)$ is no longer known. The two unknowns $C$ and $\eta$ can be determined with the two boundary conditions $\chi(-\infty)=0$, and $\chi(\infty)=\chi_f$. These can be written as
\begin{align}
    \cos(\chi(-\infty))=1=C+\eta\kappa_m\!
\left(l_m+\frac{2}{3}\,{\bf\Gamma}\!\left(\tfrac{1}{3}\right)\!z_0\right),&\\
    \cos(\chi(\infty))=\cos(\chi_f)=C-\eta\kappa_m\!
\left(l_m+\frac{2}{3}\,{\bf\Gamma}\!\left(\tfrac{1}{3}\right)\!z_0\right),&
\end{align}
from which it is straightforward to obtain
\begin{align}
    C&=\cos^2\!\left(\frac{\chi_f}{2}\right),\\
    \eta&=\frac{\sin^2\!\left(\frac{\chi_f}{2}\right)}{\eta\kappa_m\!
\left(l_m+\frac{2}{3}\,{\bf\Gamma}\!\left(\tfrac{1}{3}\right)\!z_0\right)}.
\end{align}

\begin{figure}
    \centering
    \includegraphics[width=0.8\linewidth]{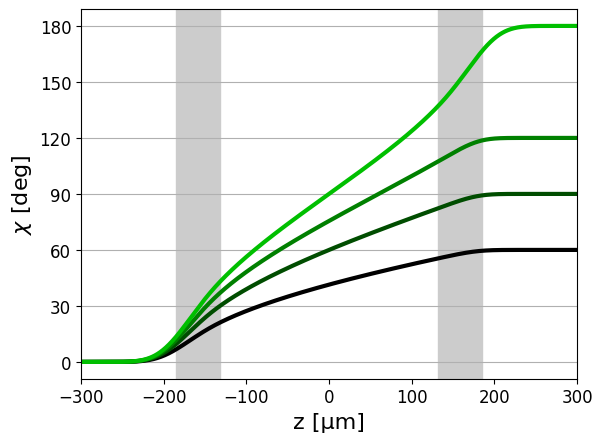}
    \caption{Coupling angle $\chi(z)$ evolution for varying values of $\chi_f$ resulting in power couplings of $25\%, 50\%, 75\%,$ and $100\%$ respectively.}
    \label{fig:fig_2SI}
\end{figure}

\end{document}